\documentclass[12pt]{article}
\usepackage{hyperref}
\usepackage{framed}
\usepackage{graphicx}
\usepackage[normalem]{ulem}
\usepackage[margin=0.75in, bottom=1in]{geometry}
\usepackage [english]{babel}
\usepackage [autostyle, english = american]{csquotes}
\MakeOuterQuote{"}

\usepackage{titlesec}

\titleclass{\subsubsubsection}{straight}[\subsection]

\newcounter{subsubsubsection}[subsubsection]
\renewcommand\thesubsubsubsection{\thesubsubsection.\arabic{subsubsubsection}}

\titleformat{\subsubsubsection}
  {\normalfont\normalsize\bfseries}{\thesubsubsubsection}{1em}{}
\titlespacing*{\subsubsubsection}
{0pt}{3.25ex plus 1ex minus .2ex}{1.5ex plus .2ex}

\makeatletter
\def\toclevel@subsubsubsection{4}
\def\l@subsubsubsection{\@dottedtocline{4}{7em}{4em}}
\makeatother

\begin{document}
\thispagestyle{plain}
\begin{titlepage}
	\begin{center}
    	\includegraphics[width=0.25\textwidth]{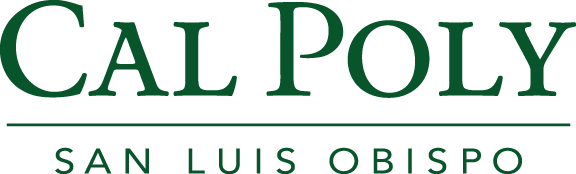}
		
        \vspace{0.9cm}
    	\Large
    	\textbf{Ethical Implications: The ACM/IEEE-CS Software Engineering Code applied to Tesla's "Autopilot" System}
    
    	\vspace{0.9cm}
    	\large
    	Kevin Vincent
    
    	\vspace{0.2cm}
    	Computer Science - Software Engineering Ethics
    
    	\vspace{0.2cm}
    	May 30, 2018
    
    	\vspace{1.8cm}
    	\textbf{Abstract}
        
    	\vspace{0.4cm}
       \end{center}
\par On October 14, 2015, Tesla Inc. an American electric car company, released the initial version of the Autopilot system \cite{blog-release}. This system promised to provide semi-autonomous driving using the existing hardware already installed on Tesla vehicles \cite{electrek}. On March 23rd, 2018, a Tesla vehicle ran into a divider at highway speed, killing the driver. This occurred under the control of the Autopilot system with no driver intervention \cite{blog-update}.
\par Critics argue that though Tesla gives drivers warnings in its owner's manual, it is ultimately unethical to release a system that is marketed as an Autopilot yet still makes grave mistakes that any human driver would not make \cite{firetruck}. Others defend Tesla by stating that their advisories are suitable and that drivers should ultimately be at fault for any mistakes of the Autopilot. This paper will scrutinize the ethical implications of Tesla's choice to develop, market, and ship a beta product that requires extensive testing. It will further analyze the implications of Tesla's aggressive advertisement of the product under the name Autopilot along with associated marketing materials. By applying the joint ACM/IEEE-CS Software Engineering Code of Ethics, this paper will show that Tesla's choices and actions during this event are inconsistent with the code and are unethical since they are responsible for adequately testing and honestly marketing their product.
\end{titlepage}

\pagenumbering{gobble}
\thispagestyle{empty}
\pagestyle{plain}
\newpage
    \onecolumn
	\tableofcontents
\newpage

\setcounter{page}{1}
\pagenumbering{arabic}
\twocolumn
\section{Facts}
Tesla released its Autopilot system on October 9th, 2014. The lead marketing materials state the following: "Full Self-Driving Hardware on All Cars. All Tesla vehicles produced in our factory, including Model 3, have the hardware needed for full self-driving capability at a safety level substantially greater than that of a human driver" \cite{tesla-autopilot}. It is currently offered as an add-on package that drivers can add either when purchasing their vehicle or even at a later date \cite{tesla-design}. Tesla's cars have driven on average over 320 million miles with one fatality which compares with 86 million miles nationally across all other vehicles \cite{blog-update}. It is designed to be aware of other cars and obstacles and stay within lines on freeways and other roadways. It is also designed to activate emergency braking if obstacles are detected. On March 23rd, 2018, Walter Huang's Tesla Model X crashed head first into a highway gore point along Highway 101 and 85 in Mountain View, California. The car at the time was under the control of Autopilot; however, it did not attempt to slow down and the driver of the vehicle died as a result of injuries sustained in the crash \cite{blog-update}.
\vspace{0.4cm}

\section{Question}
Was Tesla's response in the Walter Huang Tesla crash case ethically justifiable, and was it ethical for Tesla to release and market Autopilot as self-driving technology despite only providing warnings and disclaimers to consumers?
\vspace{0.4cm}

\section{Social Implications}
Self-driving cars and advanced driver technology are quickly becoming important additions to modern vehicles \cite{consumer_reports_2018}. It is not outlandish to imagine a near future when most if not all of our daily drives are controlled by software. Since corporations are responsible for the development, testing, release, and iteration of this software, there is a question as to whether they are ultimately responsible for mistakes the system makes. Analyzing the ethical obligations of companies like Tesla could pave a road to finally answering this question and setting precedents for future legal regulation.
\subsection{Public Safety}
Self-driving vehicles promise to increase safety and reduce fatalities at astounding levels. The potential of comprehensive sensor suites and computing in vehicles is promised to reduce accidents since over 90\% of accidents stem from driver errors \cite{nhtsa_drivers}. By eliminating the driver, a whole class of accidents, such as drunk driving, texting while driving, not seeing traffic lights, etc., can be eradicated. The social implication of this would be a massive "900,000 lives saved per year" \cite{blog-update}.
\subsection{AI and Responsibility}
As artificial intelligence systems are introduced as replacements for deterministic algorithms, particularly in self-driving vehicles, a question arises of who we hold accountable: the driver or the designer \cite{slate_responsibility}. There is also a deeper question of whether it is morally right to hand over decisions involving life and death to AI systems whose results are unpredictable? Afterward, when these systems do make mistakes, where do the law and society place the blame? Ultimately, the right answer to these questions is the one that will save the most lives.
\subsection{Automobile economy}
Self-driving vehicles will have a major impact on the automobile economy and vehicle ownership. In the future "application and content revenue generated by mobility-as-a-service will supplant the value of vehicle sales" \cite{economy_impact}. This is best exemplified by the success of current ride-sharing companies Uber and Lyft who have currently emulated the future of hailing a ride with your phone using human drivers. It has been shown that "in areas where Uber, Lyft, and other on-demand ride services operate, consumers may buy fewer cars and even take fewer trips" \cite{uber_car_owner}. As services like Uber move to self-driving vehicles, vehicle ownership and sales will further decrease leading to a major shift in today's automotive economies \cite{self-driving-ubers}.

\section{Literature Review}
\subsection{Tesla: An Update on Last Week's Accident}
Tesla's post on March 30, 2018, stated that though Autopilot was engaged, the driver ignored warnings and should have been paying attention enough to avoid the accident. Since "the driver had about five seconds and 150 meters of unobstructed view of the concrete divider with the crushed crash attenuator, but the vehicle logs show that no action was taken". Tesla believes that "no one knows about the accidents that didn't happen, only the ones that did. The consequences of the public not using Autopilot, because of an inaccurate belief that it is less safe, would be extremely severe". Tesla does not believe they should be at fault because the driver was not paying attention while using Autopilot. They further believe that it would be unethical to not release Autopilot as they believe it saves more lives \cite{blog-update}.

\subsection{Tab Turner: Tesla won't be able to put crash defense on Autopilot}
Tab Turner, a lawyer who specializes in auto-defect cases states that "there's a concept in the legal profession called an attractive nuisance... These devices are much that way right now. They're all trying to sell them as a wave of the future, but putting in fine print, Don't do anything but monitor it. It's a dangerous concept". Turner believes that Tesla is at fault and that "warnings alone are never the answer to a design problem" \cite{autonews}.

\subsection{National Highway Traffic Safety Administration: Automated Vehicles for Safety}
The NHTSA argues that "there is no vehicle currently available for sale that is 'self-driving'. Every vehicle currently for sale in the United States requires the full attention of the driver at all times for safe operation." Thus Tesla's marketing of its Autopilot system as full self-driving capability is incorrect by the NHTSA. \cite{nhtsa}.

\vspace{0.4cm}

\section{How the Software Engineering Code of Ethics Applies}
According to the IEEE/ACM Software Engineering Code of Ethics (SECOE), software engineers are individuals "who contribute by \uline{direct participation} ... to the ... \uline{development} ... and \uline{testing} of \uline{software systems}" \cite{code}.\par
\subsection{Definitions}
\subsubsection{Direct participation}
Direct participation is defined as "the process during which individuals, groups, and organizations are consulted about or have the opportunity to become actively involved in a project or program of activity" \cite{def-participation}. Tesla's software engineers are consulted and have the opportunity to actively "contribut[e] to the implementation of the software system" and thus the term direct participation holds in the domain of Tesla's Autopilot system \cite{tesla-jobs}.
\subsubsection{Testing}
In the software domain, testing is "a set of processes aimed at investigating, evaluating and ascertaining the completeness and quality of computer software."\cite{def-testing} Tesla requires that software engineers in their company have "Extensive experience creating, presenting, and refining clear, compelling technical specifications ... and test plans"\cite{tesla-jobs}. As such, it is expected that engineers that work in the company follow a set of processes to evaluate the quality of their software using these test plans. As a result, Tesla's engineers participate in testing and thus are subject to the SECOE.
\subsubsection{Development}
In the field of software, development is defined as "writing a series of interrelated programming code...using a specific programming language.., which provides the functionality of the developed software" \cite{def-development}. Tesla's engineers "develop embedded firmware in C that implements the ... software architecture for Advanced Driver Assistance Systems (ADAS) and Sensors" \cite{tesla-jobs}. Thus, Tesla's engineer's write a series of code that provides the functionality of the Autopilot software. They further develop this software using a specific programming language, C. As a result, Tesla's engineer's act of writing code to build the autopilot system falls under the definition of the term development in this domain and is subject to the SECOE.
\subsubsection{Software Systems}
A software system can be defined as "a system of intercommunicating components based on software forming part of a computer system (a combination of hardware and software)" \cite{def-softwaresys}
Tesla's Autopilot system is software that fuses input from various sensors: namely, cameras, ultrasonic, and radar, to provide advanced driver assistance capabilities \cite{tesla-jobs} As such, the Autopilot system falls under the category of a Software system and is subject to the SECOE.
\subsection{Subject to SECOE?}
As shown above, Tesla and its engineers under the company's direction, wrote, developed, tested, and shipped the code and software for Autopilot. As a result, they fall under the definition of a Software Engineer according to the SECOE and
"shall adhere to the [Software Engineering] Code of Ethics and Professional Practice" \cite{code}.

\vspace{0.4cm}

\section{Analysis of relevant SECOE Tenets}
\subsection{Tenet 1.05}
\begin{framed}
\uline{Cooperate} in efforts to address matters of \uline{grave public concern} caused by software, its \uline{installation, maintenance}, support or documentation \cite{code}
\end{framed}
\subsubsection{Definitions}
\subsubsubsection{Cooperate}
Cooperate is defined as "act jointly; work toward the same end" and "assist someone or comply with their requests" \cite{def-cooperate}.
In context, Tesla's cooperation would include working jointly with the National Transportation Safety Board and other government agencies to thoroughly investigate the event. It also includes complying with legal and other requests from these agencies and acting as directed. Cooperation would also include working with agencies to release accurate information to the public.
\subsubsubsection{Grave public concern}
A grave public concern is a public concern "giving cause for alarm" \cite{def-grave}. Examples of this can include human health and wellbeing, public security, or other public matters. According to the CDC today, "Road traffic crashes are a leading cause of death in the United States and the leading cause of death for healthy U.S. citizens".\cite{cdc_crashes} In fact, each day "an estimated 3,400 people are killed globally in road traffic".\cite{cdc_crashes} As such, "the number of people killed in car accidents every year is an alarming number"\cite{vargas_crashes} and as a result road traffic-related fatalities meet the definition of a grave public concern.
\subsubsubsection{Installation, maintenance}
Installation is defined as "the process of making hardware and/or software ready for use" \cite{def-installation}. Tesla deploys its autopilot software on vehicles for use \cite{tesla-autopilot} and even installs software updates remotely \cite{tesla-software-update}.
\subsubsection{Domain Specific Rule (DSR)}
In the domain of self-driving vehicle accidents, tenet 1.05 requires that \uline{Tesla act jointly with government investigations of road traffic related human fatalities caused by software, its deployment, software updates, support or documentation.}
\subsubsection{Analysis}
\subsubsubsection{Fatalities caused by software}
On March 23rd, 2018, a Tesla vehicle crashed into a highway divider. As it was under the autopilot system, the fatality was caused by software and more specifically its deployment onto the vehicle \cite{blog-update}. In a report released by the NTSB, it was confirmed that "Autopilot was engaged ahead of the crash" and "that a navigation mistake by Autopilot contributed to Huang's death"\cite{ntsb_report_latest}. It further gives details that the Autopilot software "began a left steering movement while following a lead vehicle", "began to accelerate, reaching a speed of 70.8mph just before the crash", and that there was no "precrash braking or evasive steering movement detected"\cite{ntsb_report_latest}. Thus, it is concluded by the report that the the Autopilot software system contributed to the fatality. As a result, this incident is subject to the Domain Specific Rule.
\subsubsubsection{Act jointly}
Tesla released investigative information before it was vetted and confirmed by the National Transportation Safety Board. This is in direct disagreement with the fact that the "National Transportation Safety Board expects parties involved in their investigations to inform them of releases before making information public." Furthermore, in discussing the situation, a National Transportation Safety Board spokesman said that "the uncoordinated release of investigative information can affect how other parties work with us in the future so we take each unauthorized release seriously" \cite{ntsb_news}. According to National Transportation Safety Board, Tesla attempted to merely speculate and make assumptions about the crash before evidence could be reviewed by government agencies. By not cooperating with the authorities, Tesla puts the public and those involved in future accidents in grave public danger by releasing unconfirmed information about ongoing investigations \cite{ntsb_news}. As a result of Tesla's mishandling of investigative information and lack of coordination with government agencies, the company was removed from the investigation by the National Transportation Safety Board \cite{ntsb_removed}.
\subsubsection{Results}
The Tesla vehicle incident of March 23rd, 2018 fulfills the definition of a human fatality caused by software and its deployment. As a result, this therefore subjects the company to the Domain Specific Rule. In addition, Tesla's release of uncoordinated and vetted information in it's investigation with a government agency is in clear violation of the "act jointly" clause of the Domain Specific Rule. By violating this principle and being subject to the Domain Specific Rule, we conclude that Tesla acted unethically in the investigation as defined by the Software Engineering Code of Ethics.

\subsection{Tenet 3.10}
\begin{framed}
Ensure \uline{adequate testing}, debugging, and review of software and related documents on which they work \cite{code}.
\end{framed}
\subsubsection{Definitions}
\subsubsubsection{Adequate testing}
Adequate can be defined as "equal to some requirement; proportionate" \cite{def-adequate}. Testing is defined in the software industry as "the process of validating and verifying that a software program/application/product: meets the business and technical requirements that guided its design and development;
works as expected; and can be implemented with the same characteristics" \cite{def-software-testing}. Combining these two definitions, adequate testing can be defined in our domain as proportionately verifying that software meets technical requirements, works as expected, and can be implemented.
\subsubsection{Domain Specific Rule}
In the domain of self driving vehicles, tenet 3.10 can be written as \uline{ensure that all software and related documents are thoroughly verified and validated to meet technical requirements, before being released to vehicles.}
\subsubsection{Analysis}
\subsubsubsection{Thoroughly verified and validated to meet technical requirements}
In some software cases, it is possible to test the full set of and types of inputs. These are described as software that has an oracle, "the tester or an external mechanism
can accurately decide whether or not the output produced by a program is correct"\cite{testing-nottestable}. Software programs that deal with well formatted input and predictable outputs are testable to a great extent. On the other hand, when the input is varied and misformed, or the "right" answer is unknown, traditional software testing practices break down \cite{testing-nottestable}. \par  The wide range of possible inputs and scenarios with self-driving vehicles is impossible to cover through standard testing practices. \cite{koopman_challenges}. As such, software companies in this domain must test to the best of their ability. As a consequence, only then can the software companies certify and deploy their software in narrow situations in which it is thoroughly tested, a method known as Phased deployment \cite{koopman_challenges}. Phased deployment is a method that in which software will work only in the specific conditions in which it was tested, but then failover when the "operational scenario suddenly becomes invalidated"\cite{koopman_challenges}..  Tesla's marketing page for Autopilot claims that "a forward-facing radar with enhanced processing provides additional data about the world on a redundant wavelength that is able to see through heavy rain, fog, dust and even the car ahead" \cite{tesla-autopilot}. It is impossible to test in all possible scenarios, in which a car will find itself in heavy rain, fog, dust, etc. As a result, this claim in Autopilot's related documentation is not thoroughly verifiable and should have been removed as it is in direct contradiction with the Domain Specific Rule.
\subsubsubsection{Verification before release}
Beta products are products usually released to customers who want to help test new features, with the prior knowledge that not all of the features may work correctly \cite{beta_def}. It is common practice for software companies who make smartphone apps, web applications like Gmail, and other software tools to perform this type of release using beta products\cite{should-beta}. On the other hand, it is uncommon and unusual for automotive companies to release beta software as "every other automaker — from BMW to Cadillac — flat-out refuses to beta test driverless tech on the public"\cite{beta_testing_bad}. It is also understood that releasing beta automotive software "is far from satisfactory, when it comes to ensuring the safety of the Tesla driver as well as everyone else on the road" even if "the driver check[s] an acknowledgment box that warns them the system is in beta"\cite{beta_testing_bad}.
\par On February 23rd, 2018, Tesla began beta-testing new software that would enhance the Autopilot system \cite{elektrek_beta}. The new software featured a new update where the vehicle would "center itself within the lane" while under the Autosteer component of Autopilot \cite{abc_similarities}. This beta functionality is blamed for the crash because the vehicle centered itself between two diverging lanes and therefore drove straight into the center barrier \cite{blame-center}. As a matter of fact, Tesla routinely releases beta software, notably software in which not all features may work correctly, to users and even charges consumers to test its beta software \cite{charge-beta}. As Tesla released\cite{elektrek_beta} and continues to release beta  software\cite{tesla_beta_new} which is defined as software in which "not all features may work correctly", the company violates the "thoroughly verified and validated...before release" clause of the Domain Specific Rule.
\subsubsubsection{Conclusion}
Tesla's lack of adequate testing and documentation regarding the strict operational scenarios in which autopilot functions, violates the adequate testing clause of the Domain Specific Rule. In addition, the release of beta software to Tesla vehicle owners violates the Domain Specific Rule condition that all software must be thoroughly tested before it is released to consumers. We can conclude that by violating these two sections of the Domain Specific Rule, Tesla did not \uline{ensure that all software and related documents are thoroughly verified and validated to meet technical requirements, before being released to vehicles} and that Tesla has therefore acted in an unethical manner.

\subsection{Tenet 6.07}
\begin{framed}
Be accurate in stating the characteristics of \uline{software on which they work}, avoiding not only false claims but also claims that might reasonably be supposed to be speculative, vacuous, \uline{deceptive, misleading}, or doubtful \cite{code}.
\end{framed}
\subsubsection{Definitions}
\subsubsubsection{Software on which they work}
In our domain, "software on which they work" can be defined as the Tesla Autopilot system, since Tesla Autopilot Engineers "contribut[e] to the implementation of the software system" \cite{tesla-jobs}.
\subsubsubsection{Deceptive, Misleading}
Deceptive is defined as "giving an appearance or impression different from the true one; misleading" \cite{def-deceptive}. Deceptive practices in this domain include providing misleading marketing materials or not stating true limitations\cite{false_advertising}. Of all the "most heavily weighed factor is the advertisement's potential to injure a customer"\cite{false_advertising}.
\subsubsection{Domain Specific Rule}
In the domain of self-driving cars, tenet 6.07 requires that Tesla \uline{be accurate in stating the characteristics of the Autopilot system, avoiding not only false marketing claims but also claims that may mislead or injure customers and not state true limitations of software}.
\subsubsection{Analysis}
\subsubsubsection{Accurate stating of characteristics of the Autopilot System}
On Tesla's main marketing site for Autopilot, they provide the broad claim that "all Tesla vehicles produced in our factory, including Model 3, \textit{have the hardware needed for full self-driving capability}"\cite{tesla-autopilot}. Tesla believes they accomplished by using forward-facing radar and cameras "eight, enabling full self-driving in almost all circumstances" \cite{tesla-autopilot}. \par On the other hand, almost every other advanced self-driving system in existence today, including those from Google, Uber, and Cruise, use a much different technology, LiDAR \cite{lidar}. LiDAR is a technology that is much more expensive than cameras but provides more refined 3D maps of the world surrounding the self-driving vehicle. Tesla is attempting to build self-driving vehicles using technology that industry has set aside as being insufficient. In fact, researchers found that "The primary functional sensor gap between today’s ADAS and higher level autonomous vehicles will be filled with the addition of LiDAR, which will help to provide reliable obstacle detection and simultaneous location and mapping (SLAM)"\cite{high_level_lidar}. Tesla claims with their statement "full self-driving capability"\cite{tesla-autopilot} that their vehicle has higher level autonomous capability yet they do not use LiDAR which has been shown an necessary for autonomous capability by researchers. As a result of unreliable object detection, in recent months, Tesla vehicles have hit a parked fire truck \cite{firetruck} and a parked police cruiser \cite{policecruiser} all while under Autopilot. This is a direct cause of Tesla's choice to forgo LiDAR as radar "systems are designed to ignore static obstacles because otherwise, they couldn't work at all"\cite{firetruck}.
Tesla's statement that their system has "the hardware needed for full self-driving capability"\cite{tesla-autopilot}, while using technology deemed by researchers to be insufficient for higher level autonomous, is a clear violation of the Domain Specific Rule.
\subsubsubsection{Misleading marketing claims, state limitations}
Tesla also makes many marketing claims on their Autopilot website and throughout their car ordering process. For example, the company claims that their "system is designed to be able to conduct short and long distance trips with no action required by the person" and will "navigate urban streets (even without lane markings), manage complex intersections with traffic lights, stop signs and roundabout" \cite{tesla-autopilot}. Nevertheless, in between these statements, the company places a statement that states "it is not possible to know exactly when each element of the functionality described above will be available, as this is highly dependent on local regulatory approval" \cite{tesla-autopilot}. This statement seems to place the availability of functionality on local regulators, when in fact the technology is incapable of providing this functionality even with regulatory approval as is shown prior. This marketing page also does not include literature that states drivers must keep their hands on the wheels at all times while the car is self-driving, despite Tesla themselves stating this in other less visible areas like blog posts \cite{blog-update} and in the owners manual\cite{owners_manual}. By placing important safety information in less visible areas, Tesla may mislead customers. Furthermore, engineers within the company "did not believe the system was ready to safely control a car"\cite{engineer-belief}. However, Elon Musk, Tesla's CEO, announced its cars would have the capability of full self-driving functionality \cite{engineer-belief}. The direct contradiction of marketing materials to internal engineers' beliefs shows that the company made claims that did not state the true limitations of their software.
\subsubsection{Conclusions}
Tesla claims that their vehicle has the hardware to enable full self-driving capability even while using insufficient technology, cameras and radar instead of LiDAR, according to industry and researchers. In addition, Tesla relegates important safety information away from marketing materials and knows that its system has safety issues according to engineers. As a result, we can conclude that Tesla violated the "be accurate in stating the characteristics of the Autopilot system" clause and "avoid false marketing claims" clause of the Domain Specific Rule and thus acted unethically according to the Software Engineering Code of Ethics.

\subsection{Tenet 6.08}
\begin{framed}
\uline{Take responsibility} for detecting, \uline{correcting}, and reporting errors in software and associated documents on which they work.\cite{code}
\end{framed}
\subsubsection{Definitions}
\subsubsubsection{Take responsibility}
Responsibility is defined as "the state or fact of being responsible, answerable, or accountable for something within one's power, control, or management" \cite{def-responsibility}. As such, in the domain of software, taking responsibility can be defined as being accountable when mistakes occur in software.
\subsubsubsection{Correcting}.
Correct is defined as "to alter or adjust so as to bring to some standard or required condition" \cite{def-correct}. In the domain of software, corrections are referred to as "patching" which involves "fixing bugs that make the software run slow or not work right"\cite{patching}. When making these corrections or patches, there are "best practices" which include "making sure you do it in a timely manner"\cite{patching}.
\subsubsection{Domain Specific Rule}
In the domain of self-driving cars, Tenet 6.08 can be written as \uline{Be accountable for mistakes in detection and correction, and take responsibility for patching software in a timely manner for self-driving software}.
\subsubsection{Analysis}
\subsubsection{Patching software in a timely fashion}
Prior to the deadly crash on March 23rd, 2018, the driver of the vehicle took his Model X to a Tesla dealership multiple times according to his family. The driver complained that the vehicle "kept veering towards the same barrier on Highway 101, near Mountain View, California" \cite{blame-center}. This is the same barrier into which the vehicle ultimately crashed, killing the driver. Tesla claims that they have no records indicating that the driver complained about this or any other Autopilot issue with his vehicle to the company \cite{blame-center}. \par In addition, the company made no immediate move to correct the software to handle similar issues after an incident occurred in September 2017. Reporters from ABC7 news found footage of another incident in which a Tesla vehicle also crashed into a center divider in similar conditions (low sun angle, clear lane markings, on Autopilot) \cite{abc_similarities}. Issuing a fix at that time may have prevented the fatality in March 2018. 
Furthermore, even a week after the deadly incident, another Tesla owner recorded a video of his vehicle also swerving towards the same center divider and posted it online \cite{abc_similarities}. Tesla could have taken action to disable Autopilot in that zone for the interim, but ultimately chose not to make any modifications or adjustments to the software.
\subsubsection{Conclusions}
In conclusion, Tesla did not fix the necessary fixes with its Autopilot system in a timely manner and thereby did not immediately acknowledge fault. This is in violation of the Domain Specific Rule clause "take responsibility for patching software in a timely manner for self-driving software". Thus, the company acted in an unethical manner according to the Software Engineering Code of Ethics.

\subsection{Tenet 6.12}
\begin{framed}
\uline{Express concerns} to the people involved when \uline{significant violations of this Code} are detected unless this is impossible, counter-productive, or dangerous.\cite{code}
\end{framed}
\subsubsection{Definitions}
\subsubsubsection{Express concern}
Concern can be defined as "a feeling of worry about something, especially one that a lot of people have about an important issue" \cite{def-concern}. In the software domain, expressing concern can be defined as showing a feeling of worry about something through the software company's communications.
\subsubsubsection{Significant Violation of this Code}.
A violation can be defined as "the action of violating someone or something."\cite{def-violation}. The ACM, in their explanation of the Software Engineering Code of Ethics states that "In all these judgments concern for the health, safety and welfare of the public is primary".\cite{code} As such, In the domain of self-driving cars and the SECOE, we can define a significant violation as the action of violating the health, safety, and welfare of self-driving car users.
\subsubsection{Domain Specific Rule}
In our domain of self-driving vehicles, Tenet 6.12 can be written as \uline{show a feeling of worry through company communications to the people involved when the health, safety, and welfare of self-driving car customers is violated unless this is counter-productive}.
\subsubsection{Analysis}
\subsubsubsection{Show feeling of worry through official communication}
After the March 23, 2018 incident, Tesla released a blog post describing what they believed may have happened. The company calls out details such as the "cruise control follow distance set to minimum", "driver's hands were not detected on the wheel 6 second prior to the accident", "no action being taken by the driver", "audible warnings earlier in ride" \cite{blog-update}. In addition, the posting makes no attempt to acknowledge that the Autopilot mistook lane markings, departed the lane, and accelerated into the barrier\cite{tesla-warnings}. It also does not say whether or not Autopilot alerted the driver that it was not confident in lane markings, which it usually performs though audio warnings \cite{tesla-warnings}. Tesla's response focuses mainly on the driver, his car configuration, and other details that take attention away from the simple fact that Autopilot was active during the accident and did not do anything to prevent the accident\cite{ntsb_report_latest}. They place all blame on the driver for not paying attention, which does not show a feeling of worry towards the people involved. In the past, the company has brought up statistical safety points whenever an incident occurred, which the company even admits was seen as lacking empathy for the situation \cite{blog-update}. An analyst stated that “Tesla explicitly uses data gathered from its vehicles to protect itself, even if it means going after its own customers”\cite{tesla_criticized}. As a result, Tesla violated the "welfare of self-driving car customers" clause of the DSR and is in violation of Tenet 6.12. 
\subsubsubsection{Unless counter-productive}
Official communication expressing deeper concern was never published following the incident \cite{tesla-blog}. In addition, it is not counter-productive or dangerous to express sympathy for victims of self-driving car accidents. Some may claim that it may slow down adoption of the technology, but expressing concern will show consumers that companies are willing to make changes to self-driving software that takes human lives into its hands \cite{brand-crisis}. As a result, Tesla's response does not meet the "unless counter-productive" clause of the DSR and therefore the company is still in violation of the Domain Specific Rule.
\subsubsection{Conclusion}
Tesla's immediate communication following the incident focused primarily on reasons why the driver was at fault and did not acknowledge or express concern for the people involved. The company focused more on the faults of the driver than the faults of their Autopilot system and showed a lack of concern for the personal feelings of those involved. As such the company violated the "show a feeling of worry through company communications" clause of the Domain Specific Rule of \uline{show a feeling of worry through company communications to the people involved when the health, safety, and welfare of self-driving car customers is violated unless this is counter-productive}. Thus, the company acted unethically according to the Software Engineering Code of Ethics.
\vspace{0.4cm}

\section{Conclusion}
On March 23rd, 2018, Walter Huang's Tesla Model X crashed head first into a highway gore point along Highway 101 and 85 in Mountain View, California. The car at the time was under the control of Autopilot; however, it did not attempt to slow down and the driver of the vehicle died as a result of injuries sustained in the crash \cite{blog-update}. \par Tesla acted as a Software Engineer in its development, testing, and release of the Autopilot system and according to the SECOE "shall adhere to the [Software Engineering] Code of Ethics and Professional Practice" \cite{code}. \par By our Analysis of Tenet 1.05, we found that Tesla was unethical in its lack of cooperation with government agencies. Analysis of 3.10 and 6.07 show that Tesla's actions were unethical in that the company failed to perform adequate testing and then communicate the possible failure modes and software limitations in deceptive marketing materials. Analysis of Tenet 6.08 and 6.12 finally show that Tesla did not express significant concern for the persons involved with the incident and that they shied away from taking full responsibility for the incorrect actions of Autopilot.\par Ultimately, the analysis of these 5 tenets shows that Tesla's advertisement of the product under the name Autopilot, its failure to adequately test and correct mistakes, and its actions during and after the death of Walter Huang was unethical by the standards set forth by the Software Engineering Code of Ethics.
\vspace{0.4cm}


\newpage
\onecolumn
\Urlmuskip=0mu plus 1mu\relax
\bibliographystyle{IEEEannot}
\bibliography{bibliography}


\newpage
\end{document}